\documentclass[aps,prl,preprint]{revtex4}
\usepackage{graphicx}
\usepackage{dcolumn}
\usepackage{bm}
\usepackage{amsmath}

\begin{document}

\title{Quantum decoherence in  a supersymmetric mechanical system }

\author{T. Shi$^{1},$ Z. Song$^{1}$ and C. P. Sun$^{1,2,a,b}$ }

\address{$^{1}$ Department of Physics, Nankai University, Tianjin 300071, China\\
$^{2}$Institute of Theoretical Physics, The Chinese Academy of
Sciences, Beijing, 100080, China}

\begin{abstract}
We make a novel observation about the decoherence phenomenon of
the fermion in the Witten's supersymmetric (SUSY) quantum
mechanical model. It is shown that, when  the bosonic partner in
the SUNY model is unobservable in a certain energy scale, the
quantum coherence meant  by  the  superposition of fermion states
can not be preserved for a long time due to the quantum
decoherence induced by the overlooked boson. This refers to a
supercharge superselection rule  similar to the charge
superselection. We numerically calculate the decoherence factor
characterizing the extent of decoherence . The obtained result
shows the periodic decoherence with finite quantum jump. Some
analytic results under the harmonic approximation for the
superpotential can be obtained to confirm the numerical
calculations.

\textbf{{Pacs numbers: 11.30.Pb, 03.65.Bz, 12.60.Jv} }
\end{abstract}

\maketitle

The original idea of supersymmetry (SUSY) is rooted in the generalization of
the Poincar\'{e} symmetry for the relativistic quantum field theory\cite{WZ}%
. Though the SUSY can not finally be confirmed as the successful
way to unify the nature laws of particle physics, the SUSY has
become a major method in non-relativistic quantum physics after
its one dimensional model was built by E. Witten\cite{Witten}. Its
successful applications could be found in many fields in physics,
such as, statistical physics and atomic physics\cite{Jun}.
However, the existence of the SUSY partners of boson or fermion
predicted by the SUSY theory in high energy physics has not been
discovered experimentally so far. The popular explanation for this
fact is to emphasize that the energy scale at which  SUSY
particles may occur should be much higher than that in all current experiments\cite%
{RMP}. In this paper we will take into account the fact of
"no-finding" of SUSY partner from a "new angle". Our central idea comes from the
quantum decoherence theory% \cite{Zeh,Zurek}.

Let us recall the superselection rule that can explain why there
does not exist the long life superposition of the states of
neutron and proton though they belong to a double sector of
isospin. The basic understanding for this phenomenon is the
isospin symmetry breaking due to the electromagnetic interaction.
From the point view of quantum decoherence \cite{Zeh}, the the
different ways of interaction between the electromagnetic field
and the neutron and proton will force the electromagnetic field to
evolve into two distinguishing states respectively and then the
states of two different isospin states, the neutron and proton,
entangle with them. This entanglement will result in the adiabatic
 decoherence \cite{sun1} in a formal superposition of the state of neutron and
proton. We notice that L. H. Ford has studied similar problem for
the electron coherence with a detailed calculation\cite{ford}. As
he showed, due to the coupling of the electron to the quantized
electromagnetic field, the influences of both of photon emission
and of the electromagnetic vacuum fluctuations can lead a decrease
in the amplitude of the interference oscillations, a kind of
quantum decoherence.

Actually, in any non-trivial SUSY theory, there must exists the
coupling between the boson (fermion) and its fermion (boson)
partner. If the SUSY partner of a particle  is invisible in an
experiment, its quantum information can be lost into the subspace
for the SUSY partner. In other words, if there exists a SUSY
system in nature, the observed boson (or fermion) could not be
preserved in a pure state (a quantum superposition) for a long
time since its unobservable SUSY partner can induce its quantum
decoherence. This argument shows that  the quantum decoherence can
result in a  superselection rule, which forbid some state
superpositions of the SUSY  particles.

Now we demonstrate the above arguments by making use the simplest,
but non trivial SUSY quantum mechanical model - the $N=2$ SUSY
quantum system introduced by E. Witten\cite{Witten}. The Witten's
SUSY model for $N=2$ is defined with two super-charges
\begin{eqnarray}
Q_{1} &=&\frac{1}{\sqrt{2}}\left( \frac{P}{\sqrt{2m}}\sigma _{1}+W(x)\sigma
_{_{2}}\right)  \\
Q_{2} &=&\frac{1}{\sqrt{2}}\left( \frac{P}{\sqrt{2m}}\sigma _{2}+W(x)\sigma
_{_{1}}\right)   \notag
\end{eqnarray}%
on the Hilbert space $H=L^{2}(R)\otimes C^{2}$. Here $L^{2}(R)$ is the space
of real-valued function and $C^{2}$ is the spin space; $\left\{ \sigma _{j}\
|j=1,2,3\right\} $ are the Pauli matrices. The real function $W(x)$ of the
Cartesian coordinate $x$ is customarily called SUSY potential. With these
notations, the SUSY Hamiltonian
\begin{equation}
H=2Q_{1}^{2}=2Q_{2}^{2}
\end{equation}%
can be explicitly written as
\begin{equation}
H=\frac{P^{2}}{2m}+W^{2}(x)+\frac{\hbar }{\sqrt{2m}}W^{\prime }(x)\sigma
_{3}.
\end{equation}%
This Hamiltonian can also be expressed in the block-diagonal form in the $%
\sigma _{3}$-picture
\begin{equation}
H=\left(
\begin{array}{cc}
H_{+} & 0 \\
0 & H_{-}%
\end{array}%
\right) .
\end{equation}%
Here, the Hamiltonian in the diagonal blocks are
\begin{equation}
H_{\pm }=\frac{P^{2}}{2m}+W^{2}(x)\pm \frac{\hbar }{\sqrt{2m}}W^{\prime }(x)
\end{equation}%
which acting on the spatial space $L^{2}(R).$ In fact, in this model, we can
understand the above Witten model as a composite system with coupling boson
and fermion. Let us define the boson creation operated
\begin{equation}
b^{\dag }=\sqrt{\frac{m\omega }{2\hbar }}\left( x-\frac{i}{m\omega }p\right)
\end{equation}%
and the fermion creation operator
\begin{equation}
f^{\dag }=\sigma _{+}=\frac{1}{2}\left( \sigma _{1}+i\sigma _{2}\right)
\end{equation}%
where $\omega $ is a constant determined by the function $W(x).$ Then the
above Hamiltonian can be written in the form
\begin{eqnarray}
H &=&\hbar \omega \left( b^{\dag }b+\frac{1}{2}\right) +W^{2}\left( \sqrt{%
\frac{\hbar }{2m\omega }}\left( b^{\dag }+b\right) \right)   \notag \\
&&-\frac{\hbar \omega }{4}\left( b^{\dag }+b\right) ^{2}+\frac{\hbar }{\sqrt{%
2m}}W^{\prime }\left( \sqrt{\frac{\hbar }{2m\omega }}\left( b^{\dag
}+b\right) \right) (2f^{\dag }f-1).
\end{eqnarray}

\bigskip

Based on  the quantum decoherence theory, we can make a crucial observation
about the SUSY Hamiltonian. The SUSY Hamiltonian describes a
quantum-measurement-like process\cite{Sun}, since the interaction part
\begin{equation}
H_{I}=\frac{\hbar }{\sqrt{2m}}W^{\prime }(x)\sigma _{3}
\end{equation}%
commutes with $\sigma _{3}\ ,$ but does not commutes with the
spatial part
\begin{equation}
H_{0}=\frac{P^{2}}{2m}+W^{2}(x)
\end{equation}%
Thus the SUSY model  a Stern-Gerlach type quantum measurement about the
fermion variable. In other word, if the system is initially prepared in the
different $\sigma _{3}$-eigen states
\begin{equation}
\left\vert +\right\rangle =\left[
\begin{array}{c}
1 \\
0%
\end{array}%
\right] ,\text{ \ \ }\left\vert -\right\rangle =\left[
\begin{array}{c}
0 \\
1%
\end{array}%
\right]
\end{equation}%
the spin part acts on the spatial part with with different "forces" depicted
by the different effective superpotentials\cite{sun1}
\begin{equation}
V_{\pm }(x)=W^{2}(x)\pm \frac{\hbar }{\sqrt{2m}}W^{\prime }(x).
\end{equation}%
These actions will create the quantum entanglement between the spin part and
the spatial part. Actually, the SUSY coupling $H_{I}$ dose not change the
internal energy of the fermion part due to $[H_{I},\sigma _{3}]=0,$but the
non-zero commutators  $[H_{I},H_{0}]\neq 0$ implies that the spatial part
can record the information of fermion with the differnt forces
\begin{equation}
F_{\pm }=-2W(x)W^{\prime }(x)\mp \frac{\hbar }{\sqrt{2m}}W^{\prime ^{\prime
}}(x)
\end{equation}%
This is just all story \ from  the so-called quantum nondemolation
measurement\cite{Sun}.

To demonstrate the creation of quantum entanglement, we assume the SUSY
system is initially prepared in a factorized state
\begin{equation}
\left\vert \psi (0)\right\rangle =\left( C_{+}\left\vert +\right\rangle
+C_{-}\left\vert -\right\rangle \right) \otimes \left\vert \phi \right\rangle
\end{equation}%
where $\left\vert \phi \right\rangle $ describes the initial spatial wave
packet $\phi (x)=\langle x\left\vert \phi \right\rangle $. Driven by the
SUSY Hamiltonian, this system will evolve into an entangled state
\begin{equation}
\left\vert \psi (t)\right\rangle =C_{+}\left\vert +\right\rangle \otimes
\left\vert \phi _{+}(t)\right\rangle +C_{-}\left\vert -\right\rangle \otimes
\left\vert \phi _{-}(t)\right\rangle
\end{equation}%
where
\begin{equation}
\left\vert \phi _{\pm }(t)\right\rangle =\exp \left( -iH_{\pm }t\right)
\left\vert \phi _{\pm }(0)\right\rangle
\end{equation}%
depicts the two-channel evolution of the SUSY system. However, $\left\vert
\psi (t)\right\rangle $, in usual, is not an orthogonal decomposition (or a
Schmidt decomposition) since $\left\vert \phi _{+}(t)\right\rangle $ and $%
\left\vert \phi _{-}(t)\right\rangle $ are not orthogonal with each other%
\cite{q-infor}. In the context of quantum decoherence theory\cite{Sun}, we
introduced the decoherence factor
\begin{equation}
D(t)=\langle \phi _{+}(t)\left\vert \phi _{-}(t)\right\rangle
\end{equation}%
whose norm $\left\vert D(t)\right\vert $ measures the extent of
orthogonalization or the entanglement. It was even pointed out that $D(t)$
also characterize the extent of decoherence of the spin part when the
spatial part of SUSY is overlooked in certain situation.

Actually, if we discard the spatial variable of the SUSY system for some
physical consideration, we only need the reduced density matrix
\begin{equation}
\rho =Tr_{x}(\left\vert \psi (t)\right\rangle \left\langle \psi
(t)\right\vert )=\left\vert C_{+}\right\vert ^{2}\left\vert +\right\rangle
\left\langle +\right\vert +\left\vert C_{-}\right\vert ^{2}\left\vert
-\right\rangle \left\langle -\right\vert +(C_{+}C_{-}^{\ast }\left\vert
+\right\rangle \left\langle -\right\vert D(t)+h.c)
\end{equation}%
to deal with the physical processes concerning the spin variable only.
Obviously, when $D(t)=0,$ the spin part of the SUSY system lose its all
quantum coherence into the spatial part and then the reduced density matrix
becomes of diagonal type, i.e.,
\begin{equation}
\rho \rightarrow \rho _{dig}=\left\vert C_{+}\right\vert ^{2}\left\vert
+\right\rangle \left\langle +\right\vert +\left\vert C_{-}\right\vert
^{2}\left\vert -\right\rangle \left\langle -\right\vert .
\end{equation}%
It is very similar to the description of double-slit interference with
"which-way" detection. If one can precisely determine which slit the
particle will go through, the interference pattern disappears.

From the above arguments it can be concluded that, if we consider the spin
1/2 variable of the SUSY system as the fermionic degree of freedom, while
the spatial parts as the bosonic degree of freedom, the quantum decoherence
of the fermion part can be induced by overlooking the motion of the bosonic
part. This implies that if we can not measure both the bosonic degrees of
freedom in an experiment, the observed part ( fermion) can not preserve its
quantum coherence since the bosonic part act as an environment to monitor
the Fermion system.

To quantitatively analyze the extent of decoherence of fermion in the SUSY
system, we go the details in the calculation of the decoherence factor. Let
us first to test the calculation in analytical way with the harmonic
approximation. The superpotential $V_{\pm }(x)$ can be expanded in the terms
of $x-x_{0}$ up to second order around the two equilibrium points $%
x_{0}^{\pm }$ satisfying $V_{\pm }^{^{\prime }}(x_{0}^{\pm })=0$ ( Fig. 1).

\vspace*{-1cm}
\begin{figure}[tbp]
\hspace{24pt}\includegraphics[width=10cm,height=15cm]{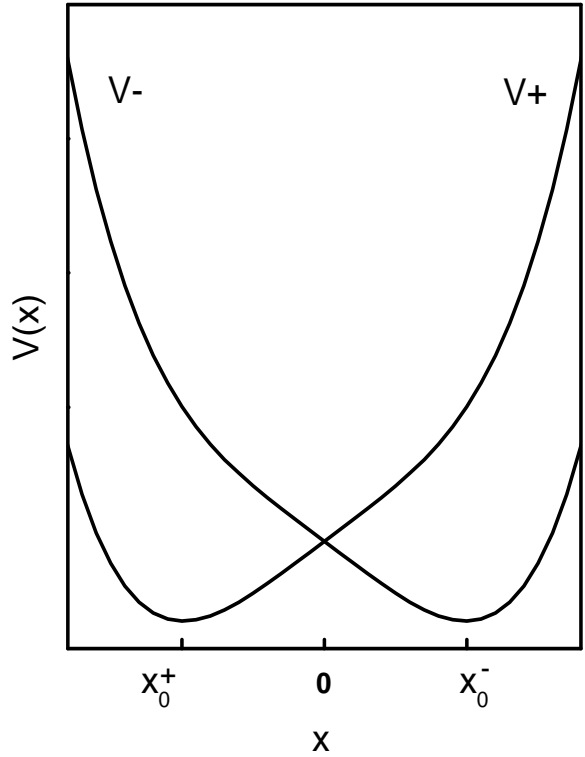} \vspace*{%
-0.0cm}
\caption{The harmonic approximation is taken in the SUSY system. The
superpotential $V_{\pm }(x)$ is expanded in the terms of $x-x_{0}$ up to
second order around the two equilibrium points $x_{0}^{\pm }$.}
\label{Fig.1}
\end{figure}
The result is just the harmonic approximation
\begin{equation}
V_{\pm }(x)=V_{\pm }(x_{0}^{\pm })+\frac{1}{2}m(\omega _{0}^{\pm
})^{2}(x-x_{0}^{\pm })^{2}
\end{equation}%
where
\begin{equation}
m(\omega _{0}^{\pm })^{2}=2W^{^{\prime }2}\left( x_{0}^{\pm }\right) \mp 4%
\frac{\sqrt{2m}}{\hbar }W^{2}\left( x_{0}^{\pm }\right) W^{^{\prime }}\left(
x_{0}^{\pm }\right) \pm \frac{\hbar }{\sqrt{2m}}W^{^{\prime \prime \prime
}}(x_{0}^{\pm }).
\end{equation}%
Furthermore,  we can write down the Hamiltonian for the two diagonal blocks
in terms of the creation and annihilation operators as
\begin{equation}
H_{\pm }=\hbar \omega _{0}^{\pm }b_{\pm }^{\dag }b_{\pm }+f_{\pm }\left(
b_{\pm }^{\dag }+b_{\pm }\right) +E_{0}^{\pm }
\end{equation}%
where
\begin{eqnarray}
f_{\pm } &=&-m(\omega _{0}^{\pm })^{2}x_{0}^{\pm }\sqrt{\frac{\hbar }{%
2m\omega _{0}^{\pm }}}  \notag \\
E_{0}^{\pm } &=&V_{\pm }(x_{0}^{\pm })+\frac{1}{2}m(\omega _{0}^{\pm
})^{2}x_{0}^{\pm 2}.
\end{eqnarray}%
Obviously, the above Hamiltonian depict two forced harmonic oscillators with
different external forces
\begin{equation}
F=m(\omega _{0}^{\pm })^{2}x_{0}^{\pm }
\end{equation}%
They will drive the two blocks evolve along two difference trajectories\cite%
{sun1}. The decoherence factor depends on the overlap of the two wave
packets along with two trajectories. The numerical calculation of the
overlapping integral shows the phenomenon of quantum jump in the progressive
decohrernce. The quantum coherence will revive in periodic time domain. This
is substantially due to the reversibility of the Schrodinger equation for
the quantum system with finite degrees of freedom. This  progressive
decohrernce phenomenon was found even theoretically in reference \cite{Sun}
in 1993, and the possibility of implementing its observation in cavity QED
experiment was also pointed out by us in ref.\cite{sun3}. In 1997 it was
also independently discussed \cite{brune} with another cavity QED setup.

Under the harmonic approximation we can obtain the results analytically. We
explicitly solve the Schrodinger equations of the evolution operators $%
U_{\pm }\left( t\right) $ in the interaction picture with the Wei-Norman
Ansatz
\begin{equation}
U_{\pm }\left( t\right) =e^{ip_{\pm }\left( t\right) }\exp e^{\alpha _{\pm
}\left( t\right) b_{\pm }^{\dag }}e^{\beta _{\pm }\left( t\right) b_{\pm
}}=e^{iQ_{\pm }\left( t\right) }D\left[ A_{\pm }\left( t\right) \right] .
\end{equation}%
where
\begin{equation}
D\left[ A_{\pm }\right] =\exp \left[ A_{\pm }^{\dag }b_{\pm }-A_{\pm }b_{\pm
}^{\dag }\right]
\end{equation}%
are the displacement operators generating the coherent states $\left\vert
\alpha _{\pm }\left( t\right) \right\rangle =D\left[ A_{\pm }\left( t\right) %
\right] \left\vert 0\right\rangle $ and%
\begin{eqnarray}
iQ_{\pm }\left( t\right)  &=&ip_{\pm }-\frac{1}{2}\left\vert A_{\pm }\left(
t\right) \right\vert ^{2},  \notag \\
A_{\pm }\left( t\right)  &=&-\alpha _{\pm }\left( t\right) =\frac{g_{\pm }}{%
\omega _{0}^{\pm }}\left[ \exp \left( i\omega _{0}^{\pm }t\right) -1\right] .
\end{eqnarray}%
The time -dependent coefficients $p_{\pm }\left( t\right) $, $\alpha _{\pm
}\left( t\right) $ and $\beta _{\pm }\left( t\right) $ can be determined by
solving the differential equations about them exactly
\begin{eqnarray}
ip_{\pm } &=&i\left( \frac{g_{\pm }^{2}}{\omega _{0}^{\pm }}-\frac{%
E_{0}^{\pm }}{\hbar }\right) t+\frac{g_{\pm }^{2}}{\omega _{0}^{\pm 2}}\left[
\exp \left( -i\omega _{0}^{\pm }t\right) -1\right]   \notag \\
\alpha _{\pm }\left( t\right)  &=&-\frac{g_{\pm }}{\omega _{0}^{\pm }}\left[
\exp \left( i\omega _{0}^{\pm }t\right) -1\right] ,  \notag \\
\beta _{\pm }\left( t\right)  &=&\frac{g_{\pm }}{\omega _{0}^{\pm }}\left[
\exp \left( -i\omega _{0}^{\pm }t\right) -1\right] ,
\end{eqnarray}%
where $g_{\pm }=f_{\pm }/\hbar .$. In the position reprsetation  they just
are the two wave packets
\begin{equation}
\Psi _{\pm }\left( t,x\right) =\left\langle x\right\vert \alpha _{\pm
}\left( t\right) \rangle
\end{equation}%
which is illustrated in Fig2.

\vspace*{-0.0cm}
\begin{figure}[tbp]
\hspace{24pt}\includegraphics[width=10cm,height=15cm]{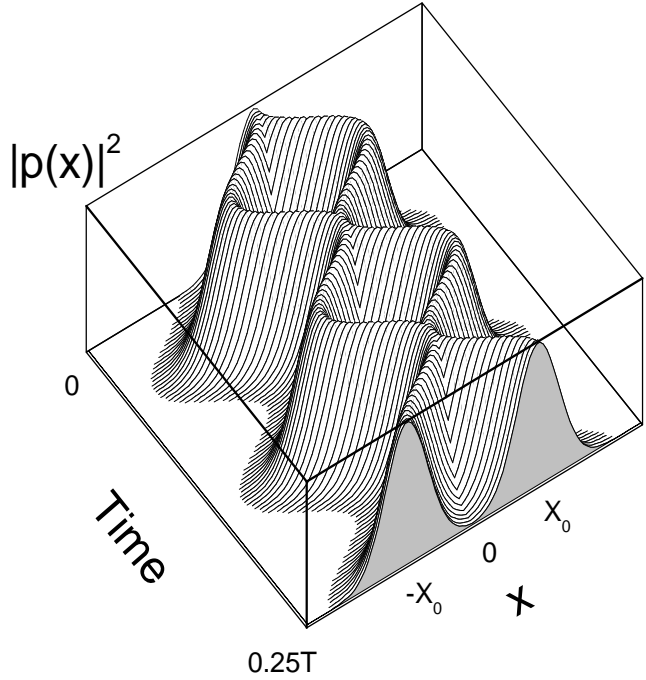}
\vspace*{-0.0cm}
\caption{The schematic illustration of the wave packet evolves in the SUSY
potential.}
\label{Fig.2}
\end{figure}

By a straightforward calculation we can analytically obtain the decoherence
factor as
\begin{eqnarray}
D\left( t\right)  &=&\left\langle \Psi _{+}\left( t\right) \right\vert \Psi
_{-}\left( t\right) \rangle =\left\langle 0\right\vert U_{+}^{\dag }\left(
t\right) U_{-}\left( t\right) \left\vert 0\right\rangle =e^{iQ_{-}\left(
t\right) -iQ_{+}\left( t\right) }\times   \notag \\
&&\exp \left[ -\frac{1}{2}\left( A_{+}W-A_{+}^{\dag }Y\right) \left(
A_{+}^{\dag }X-A_{+}V\right) \right. \left. -\left( A_{+}W-A_{+}^{\dag
}Y\right) A_{-}-\frac{1}{2}\left\vert A_{-}\left( t\right) \right\vert ^{2}%
\right]
\end{eqnarray}%
where
\begin{eqnarray}
V &=&\frac{1}{2}\sqrt{\frac{\omega _{0}^{+}}{\omega _{0}^{-}}}+\frac{im}{2}%
\sqrt{\omega _{0}^{+}\omega _{0}^{-}},  \notag \\
W &=&\frac{1}{2}\sqrt{\frac{\omega _{0}^{+}}{\omega _{0}^{-}}}-\frac{im}{2}%
\sqrt{\omega _{0}^{+}\omega _{0}^{-}},  \notag \\
X &=&\frac{-i}{2m}\sqrt{\frac{1}{\omega _{0}^{+}\omega _{0}^{-}}}-\frac{1}{2}%
\sqrt{\frac{\omega _{0}^{-}}{\omega _{0}^{+}}},  \notag \\
Y &=&\frac{-i}{2m}\sqrt{\frac{1}{\omega _{0}^{+}\omega _{0}^{-}}}+\frac{1}{2}%
\sqrt{\frac{\omega _{0}^{-}}{\omega _{0}^{+}}}.
\end{eqnarray}

To illustrate the above formulism visually, a simple example%
\begin{equation}
W=\frac{1}{\sqrt{2}}Cx^{2}
\end{equation}%
is chosen here. In the following we take $m=\hbar =1$ for simplicity. It is
easy to find that
\begin{equation}
x_{0}^{\pm }=\pm \sqrt[3]{\frac{1}{2C}},\omega _{0}^{\pm }=\omega _{0}=6C^{2}%
\sqrt[3]{\frac{1}{4C^{2}}}
\end{equation}%
and
\begin{equation}
D\left( t\right) =\exp \left( -\frac{8g_{\pm }^{2}}{\omega _{0}^{2}}\sin ^{2}%
\frac{\omega _{0}t}{2}\right) =\exp \left( -4\omega _{0}x_{0}^{2}\sin ^{2}%
\frac{\omega _{0}t}{2}\right) .
\end{equation}%
This calculations under the harmonic approximation confirm the numerical
prediction about the quantum jump in the progressive decoherence, which is
plotted in Fig. 3 for $c=0.1,0.5,1.0.$

\vspace*{-0.0cm}
\begin{figure}[tbp]
\hspace{24pt}\includegraphics[width=10cm,height=15cm]{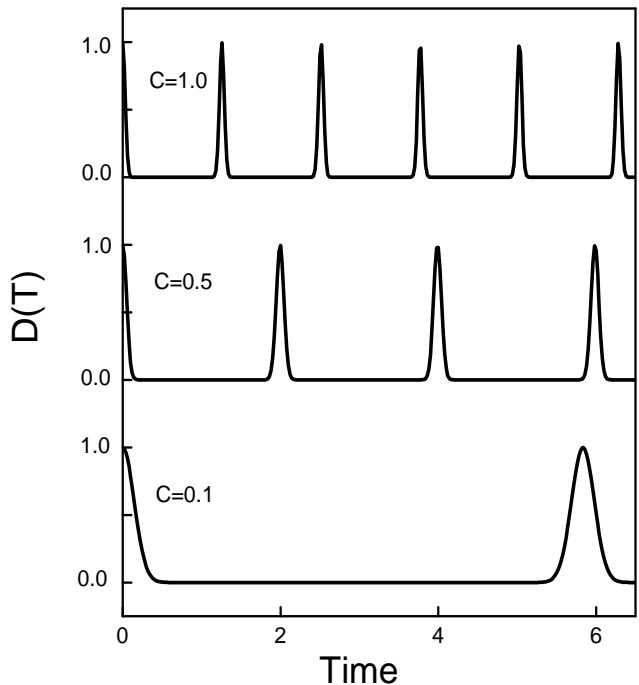} \vspace*{%
-0.0cm}
\caption{The decoherence factors as function of time for $c = 0.1, 0.5, and
1.0$.}
\label{Fig.3}
\end{figure}

Finally we conclude this paper with two remarks. 1. For the quantum
decoherence phenomenon in high energy regime, we even studied the quantum
decoherence in neutrino flavor oscillation caused by an environment
surrounding neutrinos\cite{sz}. In this way, the Ellis, Hagelin, Nanopoulos
and Srednicki (EHNS) mechanism\cite{elis} for solving the solar neutrino
problem can be comprehended in a framework of the ordinary quantum
mechanics. Because the two kinds of neutrinos have the different interaction
the weak- coupling \ environment, a microscopic model was proposed to
describe the transition of two neutrino system from a pure state to a mixed
state. It gives the modified formula of survival probability of neutrino
oscillation with two additional time-dependent parameters. For specified
environments, this result shows that the oscillating phenomena of neutrino
still exist even without a mass difference in free neutrino. 2. As has been
mentioned in the beginning of this paper, the SUSY was originally invoked to
particle physics in search of the gainful unification of nature laws with
space-time and internal symmetries. The self-consistent physics in usual
works well in the four dimensional quantum field theory. Our arguments about
the decoherence induced by the invisible SUSY partners seem to be limited to
the non-relativistic regime and thus are not able to resolve the final
problem in SUSY theory. However, we believe that our central idea can be
generalized for the relativistic quantum field theories. In this sense, the
facts revealed in our studies in the non-relativistic can be universal. We
have to point out the possible problems in our studies that the quantitative
relations between the extent of decoherence and the SUSY coupling predict
the fundamental SUSY gauge theories.

{\ We acknowledge the support of the CNSF (grant No. 90203018), the
Knowledge Innovation Program (KIP) of the Chinese Academy of Sciences and
the National Fundamental Research Program of China (No. 001GB309310).}


\begin{thebibliography}{a}
\bibitem[a]{email} Electronic address: suncp@itp.ac.cn

\bibitem[b]{www} Internet www site: http://www.itp.ac.cn/\symbol{126}suncp

\bibitem{WZ} J. Wess and B. Zumino, Nucl. Phys. \textbf{B}70, 39(1974).

\bibitem{Witten} E.Witten, Nucl. Phys. \textbf{B}188, 513(1981); \textbf{B}%
202(1982), 253.

\bibitem{Jun} G.Junker, S\textit{upersymmetric Methods in Quantum and
Statistical Physics,} (Text and Monographs in Physics, Springer-Verlag,
Berlin, 1996) .

\bibitem{RMP} Y. Shadmi and Yu. Shirman, Rev. Mod. Phys. 72, 25-64 (2000),
refs therein.

\bibitem{Zeh} E. Joos, H.D. Zeh, C. Kiefer, D. Giulini, J. Kupsch and I.O.
Stamatescu, (eds.) \textit{Decoherence and the Appearance of a Classical
World in Quantum Theory,} (Berlin:Springer-Verlag 2003)

\bibitem{Zurek} W. Zurek, Rev. Mod. Phys. 75, 715 (2003).

\bibitem{sun1} C.P. Sun, X. F. Liu, D. L. Zhou, S. X. Yu, Phys. Rev. \textbf{%
A} 63,01211, (2001); Eur. Phys. J. \textbf{D} 13, 145 (2001).

\bibitem{ford} L. H. Ford, Phys. Rev. \textbf{D} 47, 5571-5580 (1993)

\bibitem{Sun} C.P. Sun, Phys. Rev. \textbf{A} 48, 878, (1993); Chin. J.
Phys. 4, 7(1994); in Quantum-Classical Correspondence, edited by D.H. Feng
and B.L. Hu International Press, Somerville, MA, (1997), pp. 99--106; in
Quantum Coherence and Decoherence, edited by K. Fujikawa and Y.A. Ono
Elsevier Science, Amsterdam, (1996), pp. 331--334; C.P. Sun et al. Fortschr.
Phys. 43, 585 (1995).

\bibitem{q-infor} D. Bouwmeeste, A. Ekert, and A. Zeilinger (Ed.), \textit{%
The Physics of Quantum Information} (Springer, Berlin, 2000).

\bibitem{sun3} C. P. Sun, \textit{et.al.}, Quantum Semiclassic Opt \textbf{9}%
, 119 (1997).

\bibitem{brune} M. Brune,\textit{et.al}, Phys. Rev. Lett. 77, 4887 (1996);
J. M. Raimond, M. Brune and S. Haroche, Phys. Rev. Lett 79, 1964 (1997).

\bibitem{sz} C. P. Sun, D.L.Zhou, Quantum decoherence effect and neutrino
oscillation, hep-ph/9808334,1998

\bibitem{elis} J. Ellis, J. S. Hagelin, D. V. Nanopoulos and M. Srednicki,
Nucl. Phys. B 241, 381(1984).
\end{thebibliography}
\end{document}